\documentclass[12pt]{iopart}        % NEW JOURNAL OF PHYSICS

\usepackage{citesort}
\usepackage{graphicx}

\usepackage{color}
\usepackage{txfonts}

\newcommand{\eq}[1]{Eq.~(\ref{#1})}

\newcommand{\fig}[1]{Fig.~\ref{#1}}
\newcommand{\be}[1]{\begin{equation}\label{#1}}
\newcommand{\ee}{\end{equation}}

\newcommand{\mrm}{\mathrm}

\def\epsilon{\varepsilon}
\def\f{\mrm{f}}\def\i{\mrm{i}}
{\catcode`\|=\active 
 \gdef\Braket#1{\left<\mathcode`\|"8000\let|\bravert {#1}\right>}}
\def\bravert{\egroup\,\vrule\,\bgroup}
\begin{document}
\title{Strong field dynamics with ultrashort electron wave packet replicas}
\author{Paula Rivi{\`e}re, Olaf Uhden, Ulf Saalmann and Jan M Rost}

\address{Max-Planck Institute for the Physics of Complex Systems\\ 
  N{\"o}thnitzer Str. 38, 01187 Dresden, Germany.}
\date{\today}
\begin{abstract}
We investigate theoretically electron  dynamics under a
VUV attosecond pulse train which has a controlled phase delay
with respect to an additional strong infrared laser field.  Using
the strong field approximation and the fact that the attosecond pulse
is short compared to the excited electron dynamics, we arrive at
a minimal analytical model for the 
kinetic energy distribution of the electron as well as the photon absorption
probability as a function of the phase delay between the
fields.  We analyze the dynamics in terms of electron wave packet
replicas created by the attosecond pulses.
The absorption probability
shows strong modulations as a function of the phase delay for VUV
photons of energy comparable to the binding energy of the electron,
while for higher photon energies the absorption probability does not
depend on the delay, in line with the experimental observations for helium
and argon, respectively.
\end{abstract}
\pacs{32.80.Fb, 33.20.Xx, 33.60+q, 42.50.Hz}
%32.80.Fb        Photoionization of atoms and ions
%33.20.Xx        Spectra induced by strong-field or attosecond laser irradiation
%33.60.+q        Photoelectron spectra
%42.50.Hz        Strong-field excitation of optical transitions in quantum systems; multiphoton processes; dynamic Stark shift

\maketitle
\tableofcontents

\section{Introduction}
Technological advance has made it possible to expose atoms and molecules to a
combination of attosecond pulse trains (APT) and infrared (IR) laser
pulses with an accurate control of their phase delay \cite{Paul2001}.
The photo electron spectrum of atoms in this combined light field has been studied
\cite{Johnsson2005}, as well as above threshold ionization \cite{Guyetand2005},
the latter together with high harmonic generation in the combined field
also theoretically \cite{Faria2006}, along with another quasi-analytical formulation
\cite{Yudin2008} and a fully numerical R-matrix calculation \cite{Vanderhart2008}.

While many parameter combinations are possible, a dynamically very
interesting regime emerges when the energy of the VUV photon from the
APT is comparable to the ionization potential but the IR pulse alone
(typically 780 nm wavelength) is not intense enough to ionize
the atom.  The combined action of both fields leads to a
time-dependent wave packet dynamics which is very sensitive
to the phase delay, equivalent to the carrier-envelope phase (CEP) of the IR field.
Consequently stroboscopic measurements of
interfering electron wave packets which overlap in this energy regime but
 leave in opposite directions have been demonstrated to provide
  a sensitive tool to measure phase differences
 \cite{Remetter2006}. In a subsequent experiment
it has been recently shown that the phase delay does not only influence the photo electron spectrum,
but also modulates the total ionization and absorption probability \cite{Johnsson2007}.
A solution of the one-electron time-dependent
Schr{\"o}dinger equation (TDSE)  with a pseudo potential for helium
yields excellent agreement with this experiment \cite{Johnsson2007}.
While quite generally interference of wave packets must be responsible
for the pronounced oscillatory behavior of the absorption yield as a
function of the phase delay, the exact reason  and systematics of
these oscillations is difficult to identify in a fully numerical
solution.

Here, we formulate a minimal  analytical approach.  It
elucidates  the mechanism behind the
pronounced structures  in the electron observables as a function of phase delay
 in the spirit of the ``simple man's approach''
successfully formulated for high harmonic generation in intense fields \cite{Lewenstein}.
For a recent refinement of the simple man's approach, see \cite{Smirnova}.

\section{Ionization with an APT in the presence of an IR field: Analytical approach}
%%%%%%%%%%%%%%%%%%%%%%%%%%%%%%%%%%%%%%%%%%%%%%%%%%%%%%%%%%%%%%%%
We consider an electron bound with energy $\epsilon_\i$
exposed to a train of attosecond pulses in one dimension. The
intensity is so weak that we are in the single photon regime for
each atto pulse of width $\sigma$ and central time
$t_n$. Consequently we neglect the slowly varying envelope of
the ATP and write for the  
interaction potential (in velocity gauge and rotating wave approximation)
\begin{equation}
\label{ vatto}
V_{\mrm{atto}}(t)= p\, \exp(-i\Omega t)\sum_{n=1}^{N_{\mrm{atto}}}\exp[-(t-t_{n})^{2}/(2\sigma^{2})]\,,
\end{equation}
where $p$ is the electron momentum and $\Omega$ the central frequency of the attosecond VUV pulses.
The infrared field is characterized by the vector potential 
\begin{equation}
\label{ir }
A(t) = A_{0} \cos(\omega t + \varphi)\,.
\end{equation}
It has linear polarization in the same direction as the atto
pulses and a phase delay $\varphi$ with respect to the APT. 
Its envelope $A_{0}$ is assumed  to be constant over the length of the APT.
We use atomic units throughout the paper unless stated otherwise.

Since the APT gives only rise to single-photon absorption events, 
first-order time-dependent 
perturbation theory provides an accurate description under which
an initial state $|\Psi(t_\i)\rangle$ 
evolves into the state at time $t_\f$ according to
\begin{equation}
|\Psi(t_\f)\rangle=-i\int_{t_\i}^{t_\f}{U}(t_\f,t)
{V}_{\mrm{atto}}(t){U}(t,t_\i)|\Psi(t_\i)\rangle\,dt
\label{psit}
\end{equation}
with the time evolution operators $U$
for the electron under the combined influence of the IR field
and the Coulomb potential of the ion \cite{Quere2005}. 

The key idea behind the simple man's 
simplification which permits an analytical treatment is
to consider the phases in the integral of \eq{psit}
as the dominant contributions, 
$U_{\alpha}(t'',t')=\exp(i[\Phi_{\alpha}(t'')-\Phi_{\alpha}(t')])$,
and to approximate the propagators before and after the photoabsorption differently,
$U(t,t_\i)\equiv U_\i$ and $U(t_\f,t)\equiv U_\f$. 
Then, one gets from \eq{psit}
\begin{equation}
\label{psi-phase}
\psi(t)=\sum_n \int_{t_\i}^{t_\f}\exp{(\phi_n}(t))\,dt,
\label{psitap}
\end{equation}
where the phase is given by
\begin{equation}
\phi_n(t;p,\varphi)=i[\Phi_\f(t_\f)-\Phi_\f(t)]-(t-t_{n})^{2}/(2\sigma^{2})
-i\Omega t+i[\Phi_\i(t)-\Phi_\i(t_\i)]. 
\end{equation}
Here, $\Phi_\i$ is the phase accumulated  before the VUV
absorption, where the electron remains in its initial state, 
almost unperturbed by the laser field. Hence, we may write
$\Phi_\i(t)=-\epsilon_\i t$,  and define
the energy $\epsilon_\f=\epsilon_\i+\Omega$  after the
absorption of the VUV photon at time $t'$. For $t>t'$ 
we assume that the electron dominantly feels the IR field,
described by  the corresponding classical action (or,
equivalently, quantum Volkov propagator)  
\be{volkov}
\Phi_\f(t;p,\varphi)= -\int^{t}(p+A_{0}\cos(\omega\tau{+}\varphi))^{2}/2\,d\tau. 
\ee
A wavefunction expressed as an integral over time with two phases $\Phi_{\i,\f}$, 
as in \eq{psitap}, appears also in the analytical description of
high harmonic generation, see e.\,g. Eq.~(22) in \cite{Saro00}.  
In this context, often a stationary phase approximation is invoked. This is not possible for \eq{psi-phase},
since the attosecond pulse restricts the time integration
effectively to an interval of the order of $\sigma <
2\pi/\omega$,  the period of the IR field.
However, here we may expand the phases in a 
Taylor series about the atto peak 
at time $t_{n}$. This converts \eq{psitap}  for $t_\i\to-\infty$
and $t_\f\to+\infty$ into a Gaussian integral of the form
\begin{equation}
\label{psi-gauss}
\psi^{\infty}(p,\varphi) =\sum_{n} \int_{-\infty}^{\infty}\exp\left[\sum_{k=0}^{2}F_n^{(k)}(p,\varphi)\frac{(t-t_{n})^{k}}{k!}\right]\,dt\,,
\end{equation}
where the wavefunction depends now on the asymptotic electron momentum $p$  and the phase delay $\varphi$.
The solution to \eq{psi-gauss} reads
\begin{equation}
\psi^{\infty} = \sum_n\sqrt{\frac{2\pi}{-F^{(2)}_n}}\exp
\left[F^{(0)}_n-\left(F^{(1)}_n\right)^2\Big/\left(2F^{(2)}_n\right)\right]\,.
\label{uninf}
\end{equation}
The  expressions $F^{(k)}_n$ are explicitly given in \ref{sec:TaylCoeff}.
The photo-electron spectrum $dP(p,\varphi)/dp$ and the total photo absorption probability $P(\varphi)$ read in terms of
$\psi^{\infty}(p,\varphi)$
\begin{equation}
\label{observables}
dP(p,\varphi)/dp = |\psi^{\infty}(p,\varphi)|^{2},\,\,\,\,\,\,\,\,\,\,\,P(\varphi)=\int dp |\psi^{\infty}(p,\varphi)|^{2}\,.
\label{dpdp}
\end{equation}
For simplicity, we will consider an absorption probability, normalized by the number of atto pulses $N_{\mrm{atto}}$
\begin{equation}
P^{N_{\mrm{atto}}}(\varphi)\equiv\frac{P(\varphi)}{N_{\mrm{atto}}}.
\label{pm}
\end{equation}

%%%%%%%%%%%%%%%%%%%%%%%%%%%%%%%%%%%%%%%%%%
\section{Explicit form of the replicated electron wave packet}\label{timeev}
As we will see it is  possible to factorize the replicated
electron wave packet (EWP) into
one term which depends on $N$, the number of IR cycles over which the APT extends, and $\nu$, the number of atto peaks in each IR cycle. 
Experimentally, both $\nu=1$ and $\nu=2$ have been realized \cite{Mauritsson2006}.
The total number of attosecond pulses is then $N_{\mrm{atto}}=N\nu$.

We start with the fundamental APT
with one attosecond pulse in each IR period, therefore $t_n=2\pi n/\omega$.
Any offset in time can be absorbed in  the definition of the
phase $\varphi$.  
To obtain $\psi^{\infty}_{\nu}(p,\varphi)$ from \eq{psi-gauss}
explicitly we have to evaluate the functions $F_n^{(k)}$ in
\eq{uninf} at 
times $t_n$ as detailed in appendix \ref{sec:TaylCoeff}.
Collecting all the phases from \eq{uninf} but the prefactor $(2\pi/(-F_n^{(2)}))^{1/2}$, which contributes
only logarithmically to the phases,  we may write
\begin{equation}
\label{even}
\psi^{\infty}_{1}(p,\varphi)=a_{1}
\sum_{n=0}^{N-1}e^{i2n\pi C} \chi(p,\varphi)\,,
\end{equation}
where $a_{\nu}$ is an overall phase which will not affect the observables in  \eq{observables} and
\be{Cgl}
C = (p^{2}/2+U_\mrm{p}-\epsilon_\f)/\omega\,,
\ee
with the ponderomotive potential $U_\mrm{p}=A_{0}^{2}/4$.
The wave packet $\chi$ takes the form
\begin{equation}\label{chipm}
\chi(p,\varphi)=\exp(ipx_{\varphi})
\exp\left\{-\frac{[(p+p_{\varphi})^{2}/2-\epsilon_\f]^{2}}{2\sigma_{\mrm{\epsilon}}^2(p)}(1- i\sigma^{2}\omega^{2}x_{\varphi}(p+ p_\varphi))\right\},
\end{equation}
where only real valued parameters have been used.
The two parameters
\be{quiver}
p_{\varphi}=A_{0}\cos\varphi, \qquad
x_{\varphi}=(A_{0}/\omega)\sin\varphi\,,
\ee
characterize the motion of an electron released at $t=0$ into the
IR field: It will quiver around the position
$x_\varphi+p_\varphi t$ having a drift momentum $p_\varphi$.
The wave packet $\chi$ in \eq{chipm} contains
\be{epsi}
\sigma_{\mrm{\epsilon}}(p) = \left[1/\sigma^2+\left(\sigma\omega^2(p+p_{\varphi})x_{\varphi}\right)^{2}\right]^{1/2},
\ee
which is an effective width in energy with two contributions: 
The first one is $1/\sigma^2$, the variance in energy due to the temporal width of the Gaussian attosecond pulse.
The second one accounts for the {\it change} of  $p_{\varphi}$
gained from  the IR field during the VUV photo ionization. This
change is proportional to the electric field, or to
$x_{\varphi}$.  
\begin{figure}
\centering
\includegraphics[width=0.45\textwidth]{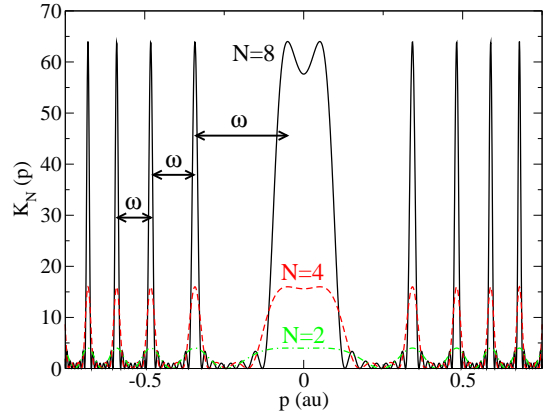}
\caption{Comb function $K_N(p)$ 
corresponding to  $N=8$ (solid),  $N=4$ (dashed) and $N=2$ (dotted-dashed)
IR cycles, for both one or two attosecond pulses per IR cycle. The
IR intensity is $I=1.3\times 10^{13}$\,W/cm$^2$
and the excess energy  $\epsilon_\f=0.144$ au.}
\label{fig:comb}
\end{figure}

Next we evaluate \eq{uninf} for two atto pulses per IR
cycle, $\nu =2$.
Apart from pulses at $t_n=2n\pi/\omega$ we have a second
sequence at $t_{n{+}1/2}=2(n{+}1/2)\pi/\omega=(2n{+}1)\pi/\omega$.
A little thought reveals that for the first sequence 
$\Phi_{\f}(t_n;p,\varphi) = \Phi_{\f}(0;p,\varphi)$,
whereas for the second one $\Phi_{\f}(t_{n{+}1/2};p,\varphi) =
\Phi_{\f}(0;-p,\varphi)$, see also appendix \ref{sec:TaylCoeff}. 
Collecting again all terms from \eq{uninf}  we can
write $\psi^{\infty}_{2}(p,\varphi)$ in the form
\begin{equation}
\label{odd-even}
\psi^{\infty}_{2}(p,\varphi)=a_{2}
\sum_{n=0}^{N-1}e^{i2n\pi C} \left(e^{-iC\pi/2}\chi(p,\varphi) + e^{iC\pi/2}\chi(-p,\varphi)\right)\,.
\end{equation}
Obviously, the sum over $n$ is the same geometric series
as in \eq{even}.  We call its absolute square the comb function, 
\begin{equation}
\label{comb}
\left|\sum_{n=0}^{N-1}e^{i2n\pi C}\right|^{2}=\frac{\sin^{2}(N\pi C)}{\sin^{2}(\pi C)}\equiv K_{N}(p)\,,
\end{equation}
and show it for increasing $N$ in \fig{fig:comb}.
From $\nu =1,2$ the general structure of the final wavefunction $\psi^{\infty}_{\nu}(p,\varphi)$ for an arbitrary number of atto pulses $\nu$ per IR cycle emerges:
It factorizes in the comb amplitude which depends
on the number of $N$ of IR cycles, and a complex wave packet
containing $\nu$ sub-packets which are created by atto pulses
during one IR cycle and  therefore  depend on the phase
difference $\varphi$ of the IR pulse and the APT.

%%%%%%%%%%%%%%%%%%%%%%%%%%%%%%%%%%%%%
\section{The photo-electron spectrum}\label{seccontrib}
\begin{figure}[t!]
\centering
\includegraphics[width=0.35\textwidth]{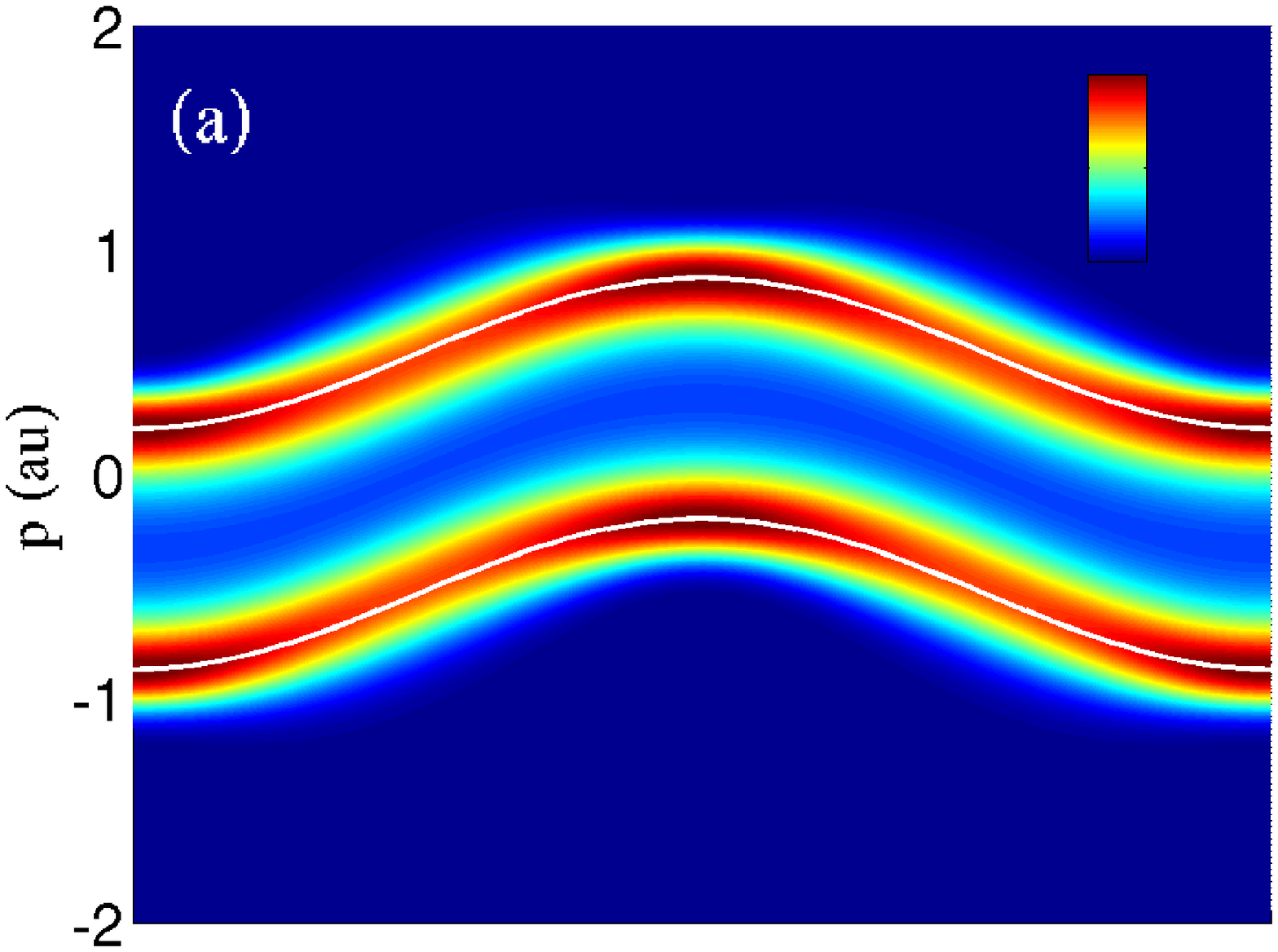}
\includegraphics[width=0.35\textwidth]{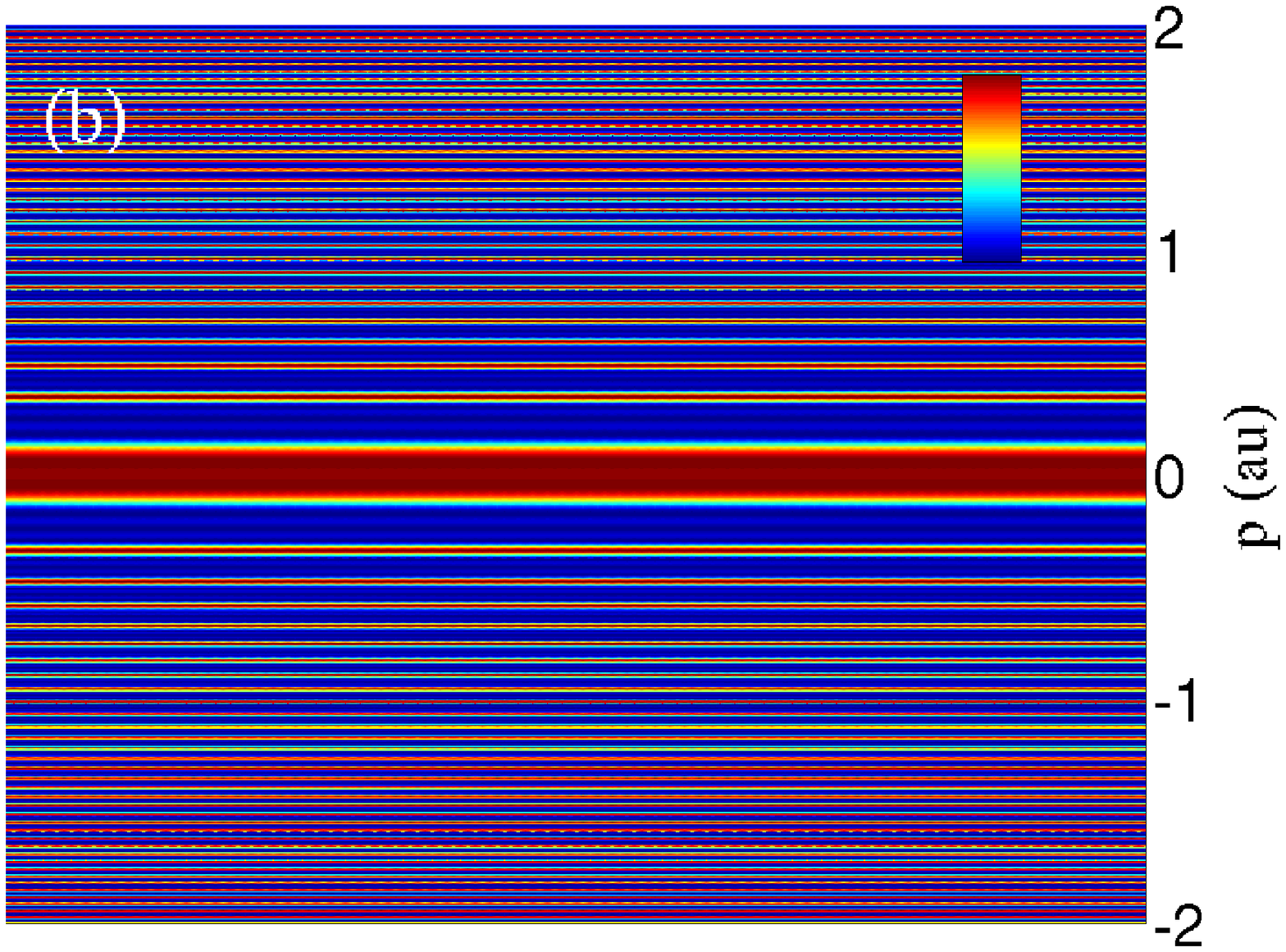}
\includegraphics[width=0.35\textwidth]{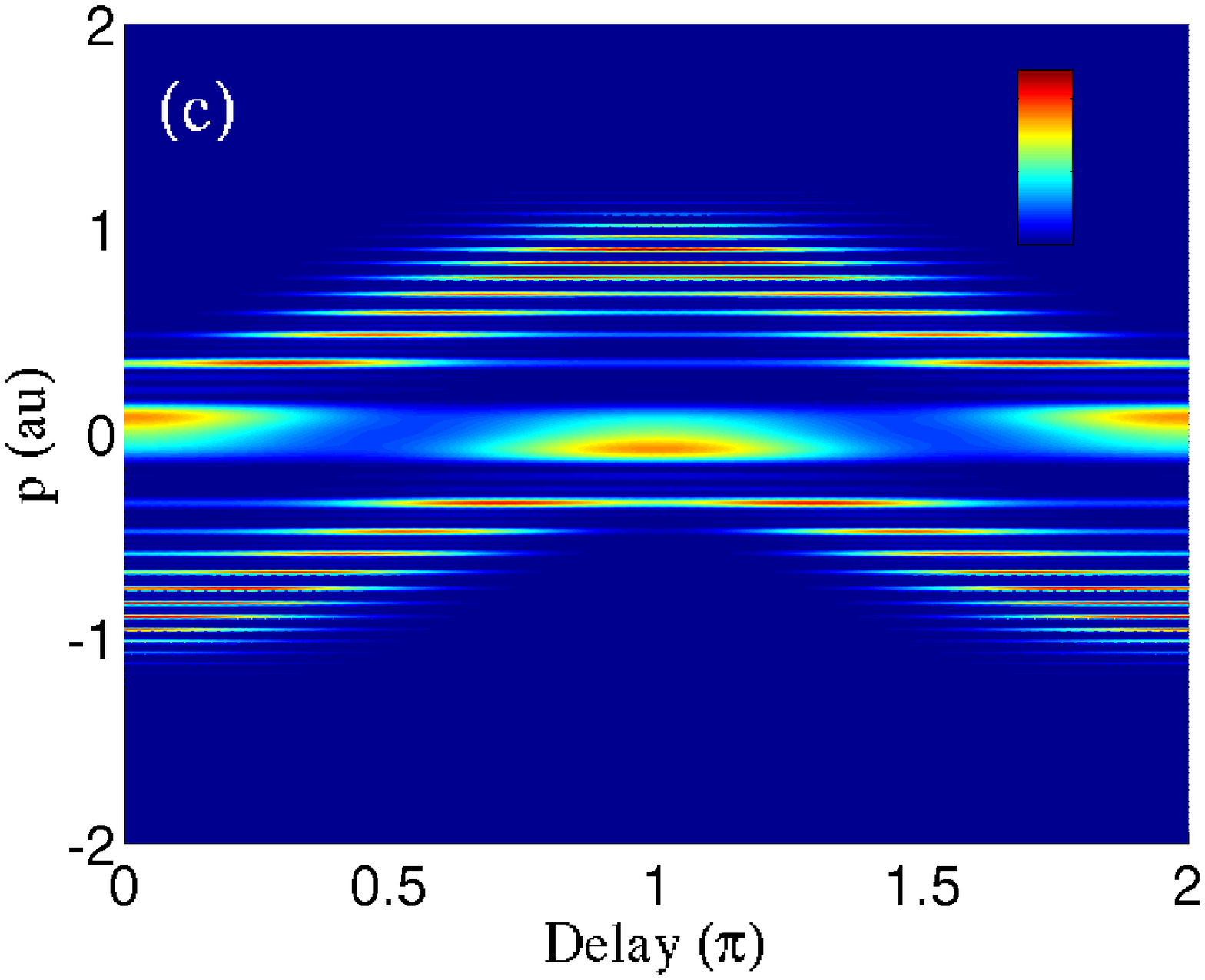}
\includegraphics[width=0.35\textwidth]{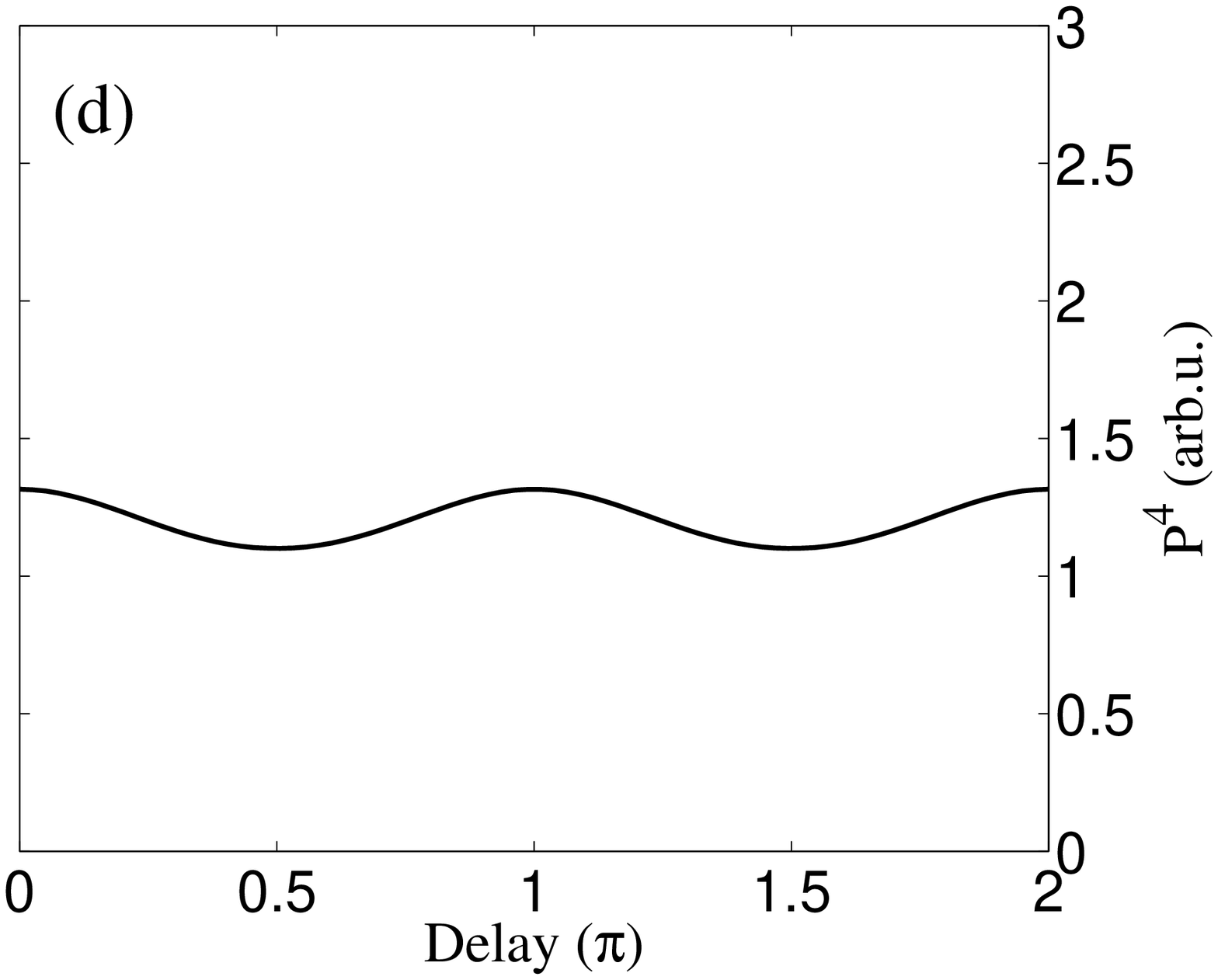}
\caption{Results for 1 atto per IR cycle, $I=1.3\times 10^{13}$\,W/cm$^2$, $\epsilon_\f=0.144$\,a.u. and $N=4$.
The attosecond pulses have a width FWHM=370 as ($\sigma=$9.186).
(a) $\Phi_{1}(p,\varphi)$. The white lines represent $\pm (2\epsilon_\f)^{1/2}-p_\varphi$.
(b) Comb function $K_{4}(p)$.
(c) Photo-electron spectrum, $|\Psi_{1}(p,\varphi)|^2$.
(d) Normalized absorption probability $P^4(\varphi)$.}
\label{fig:spectrum1}
\end{figure}
%%%%%%%%%%%%%%%%%%%%%%%%%%%%%%%%%%%%%
The product structure of the asymptotic wavefunction $\psi^{\infty}_{\nu}$ carries
over to the photo-electron momentum distribution  \eq{observables} since
\begin{equation}
dP_{\mrm{\nu}}/dp =
|\psi^{\infty}_{\nu}(p,\varphi)|^2 =  K_{N}(p){\cal X}_{\nu}(p,\varphi).
\label{cwtot}
\end{equation}
The function
$K_{N}(p)$
has maxima separated in energy by the IR frequency $\omega$ which become
sharper with increasing $N$, as can be seen in \fig{fig:comb}. Hence, $K_{N}(p)$
acts like a $comb$ in momentum for the photoelectron spectrum. The comb selects particular values of $p$ 
occurring at specific phases $\varphi$ 
from  the electron momentum distribution ${\cal X}_{\nu}(p,\varphi)$, which builds up from the $\nu$ atto pulse wave packets within one IR period.
\subsection{One atto pulse during an IR cycle}
For $\nu = 1$,  we get from \eq{chipm}
\be{X1}
{\cal X}_{1}(p,\varphi)=|\chi(p,\varphi)|^{2}=
\exp\left\{-\frac{[(p+p_{\varphi})^{2}/2-\epsilon_\f]^{2}}{\sigma_{\mrm{\epsilon}}^2(p)}\right\},
\label{x1}
\ee
The electron distribution ${\cal X}_{1}$ shown in \fig{fig:spectrum1}a has two branches centered about
$p_{\pm}=\pm\sqrt{2\epsilon_\f}-p_{\varphi}$. Each of them traces the streaking momentum $p_{\varphi}$ (\eq{quiver} and white lines in the figure) which is imprinted when the attosecond pulse excites  the electron with a phase delay $\varphi$. The width $\sigma_{\mrm{\epsilon}}$ of the branches  has maxima at $\varphi =
\frac 12\pi,\frac 32\pi$ and minima at $\varphi = 0,\pi$. The multiplication of 
${\cal X}_{1}$ (\fig{fig:spectrum1}a) with the comb $K_{4}(p)$ (\fig{fig:spectrum1}b)
gives the photo-electron momentum distribution shown in \fig{fig:spectrum1}c.
One clearly sees a preference of momenta and phase delays. The modulation in phase delays
survives upon integration over $p$ in the total absorption probability $P^{N_{\mrm{atto}}}(\varphi)$ shown in \fig{fig:spectrum1}d, with maxima at $\varphi = 0,\pi$.

\subsection{Two atto pulses during an IR cycle}
\begin{figure}[t!]
\centering
\includegraphics[width=0.35\textwidth]{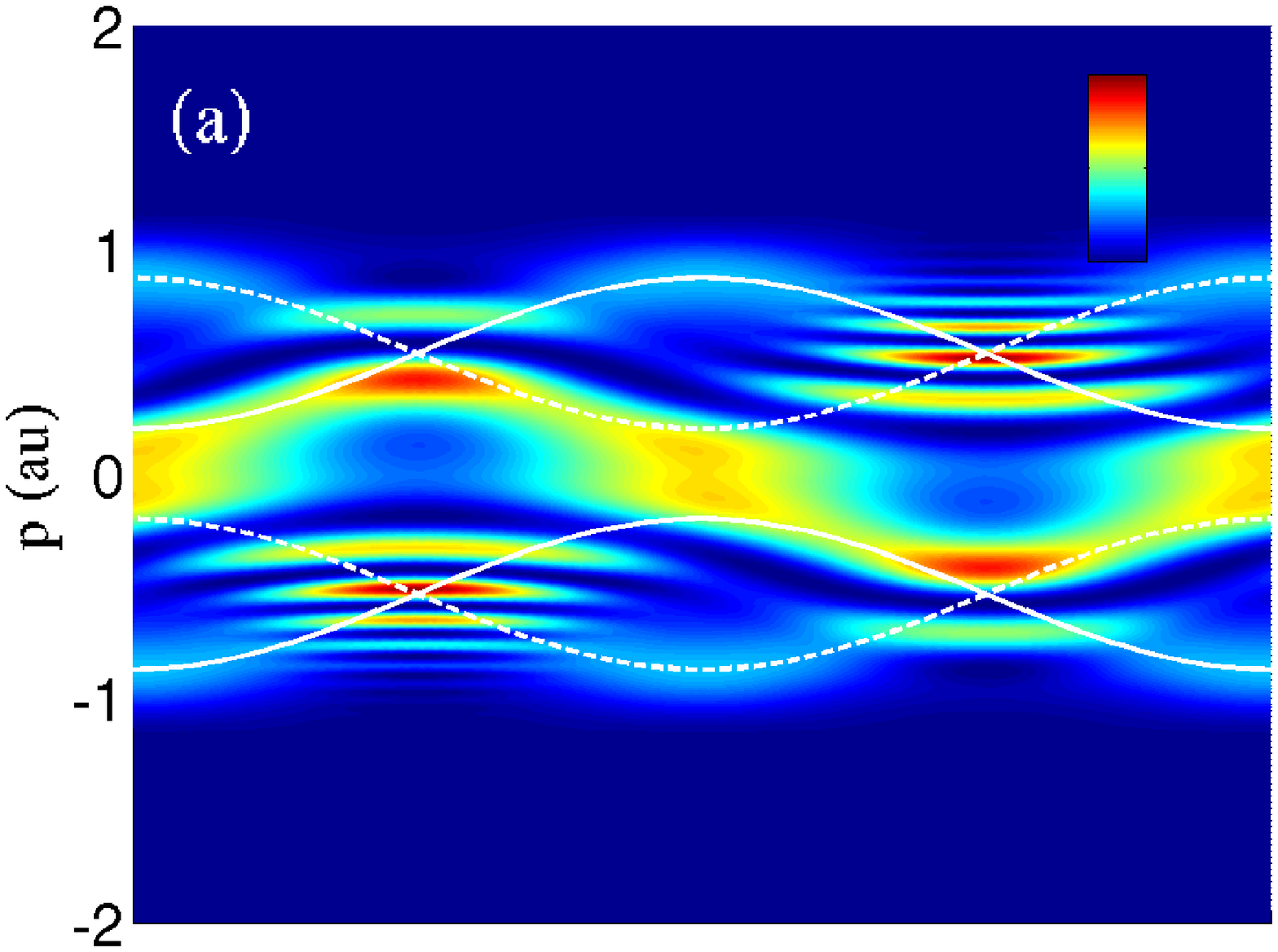}
\includegraphics[width=0.35\textwidth]{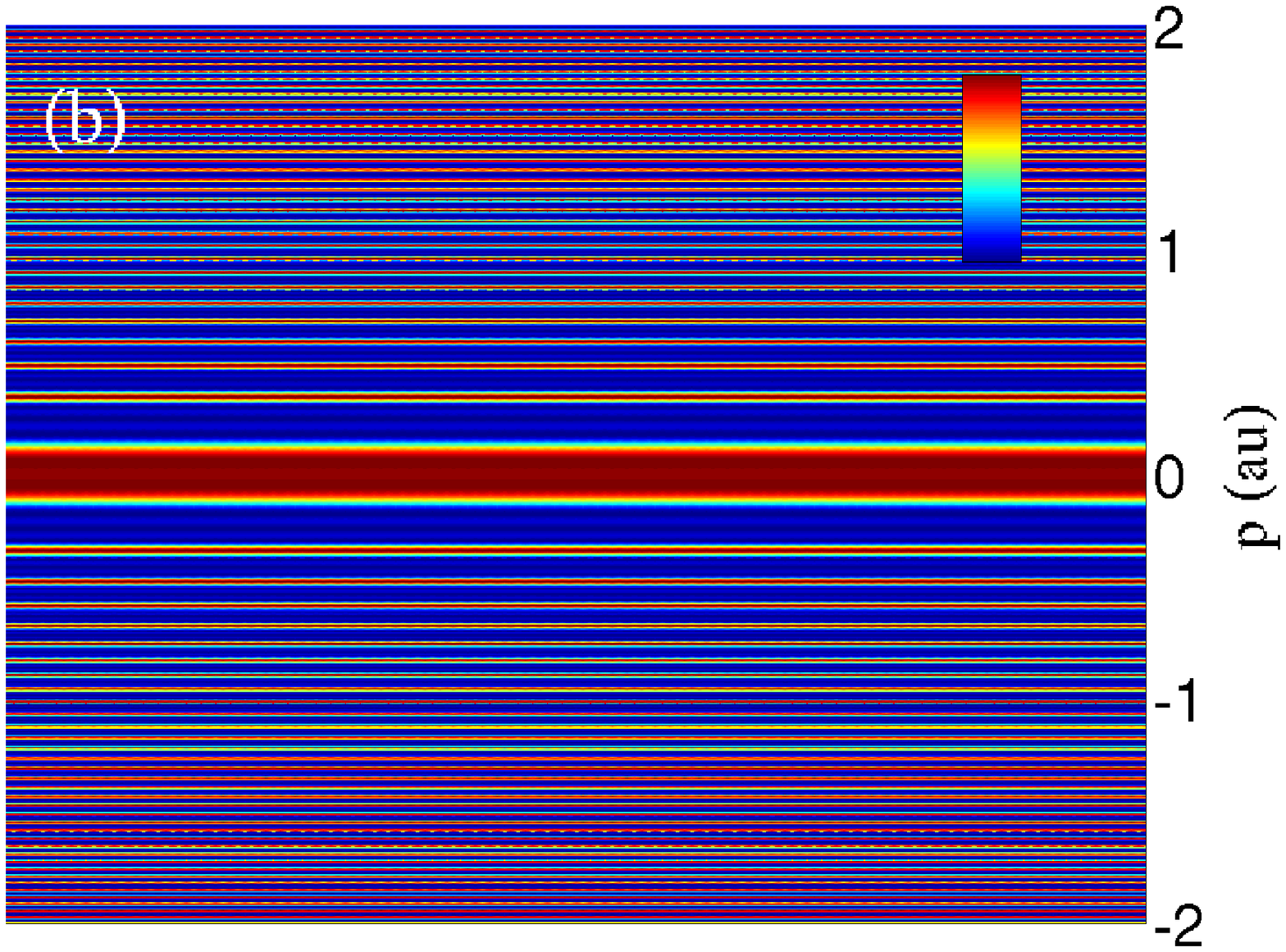}
\includegraphics[width=0.35\textwidth]{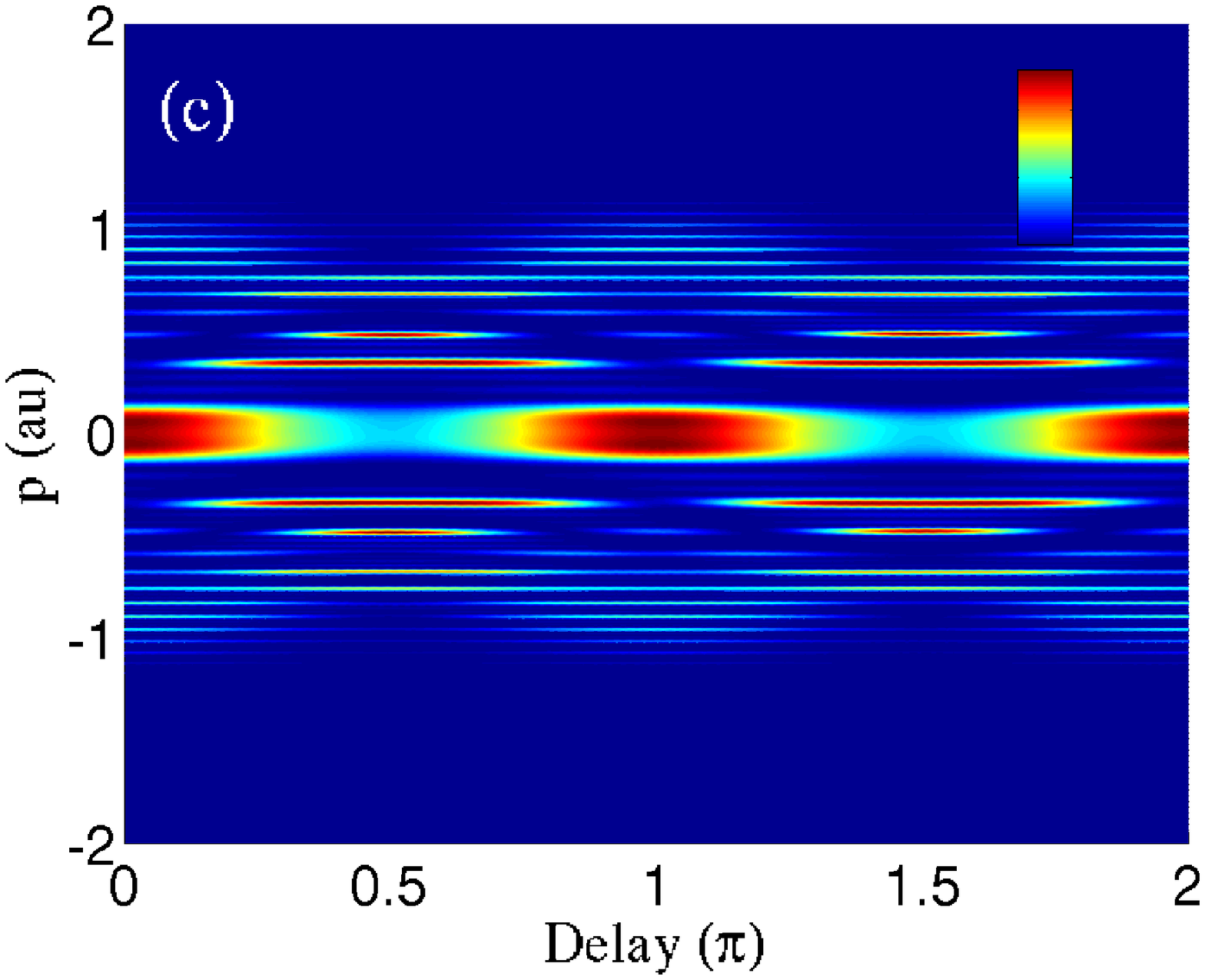}
\includegraphics[width=0.35\textwidth]{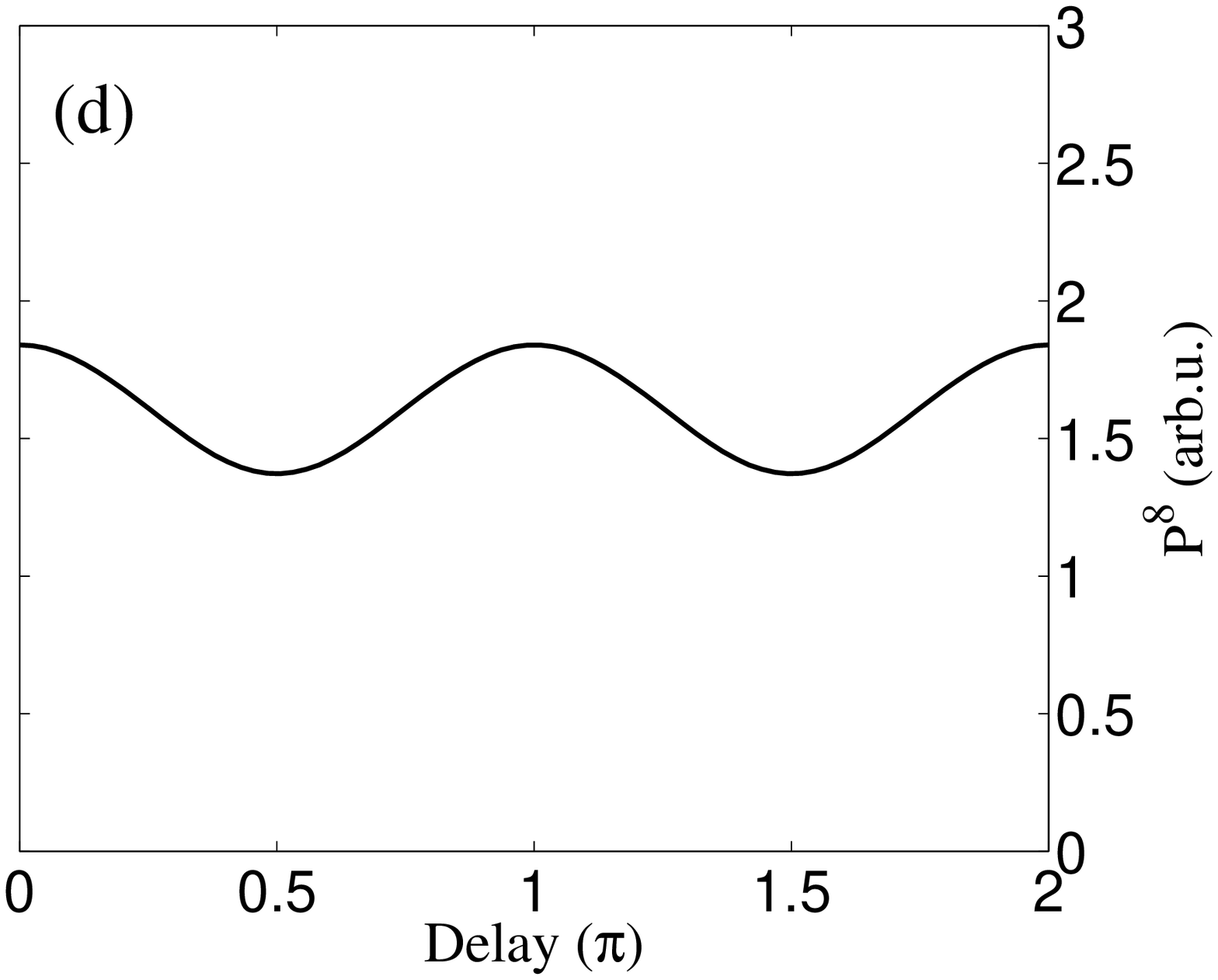}
\caption{Same as in \fig{fig:spectrum1} $(N=4)$, but for $\nu=2$ atto pulses per IR period $(N_{\mrm{atto}}=8)$. The white lines in (a) correspond
to different branches in the wave packet ${\cal X}$, see text.}
\label{fig:spectrum2}
\end{figure}
While the comb function remains the same, we have now a more complicated 
single cycle momentum distribution composed of two wave packets during each IR cycle,
\be{X2}
{\cal X}_{2}(p,\varphi)=\left|e^{-iC\pi/2}\chi(p,\varphi) + e^{iC\pi/2}\chi(-p,\varphi)\right|^{2}\,.
\ee
The basic structure with two branches for each wave packet $\chi$ is the same
as for $\nu =1$, resulting in a total of four branches at $p_{\pm}^{+}=\pm\sqrt{2\epsilon_\f}-p_{\varphi}$ and $p_{\pm}^{-}=\pm\sqrt{2\epsilon_\f}+p_{\varphi}$, indicated as white lines (solid and dashed, respectively) in \fig{fig:spectrum2}a. In addition, the  wave packets interfere leading to a rich pattern in ${\cal X}_{2}$, as can be seen in 
\fig{fig:spectrum2}a. However, again the comb $K_{4}(p)$, cf.\ \fig{fig:spectrum2}b,
selects specific momenta and phases for the photo-electron momentum  distribution \fig{fig:spectrum2}c, which produces 
a modulation in the total absorption probability (\fig{fig:spectrum2}d) 
similarly as for $\nu =1$, 
with maxima
at $\varphi = 0,\pi$.
Note, that for both $\nu = 1$ and $\nu =2$ the number of maxima
in the absorption probability $P_{\mrm{\nu}}(\varphi)$ is the same.

\begin{figure}
\centering
\includegraphics[width=0.4\textwidth]{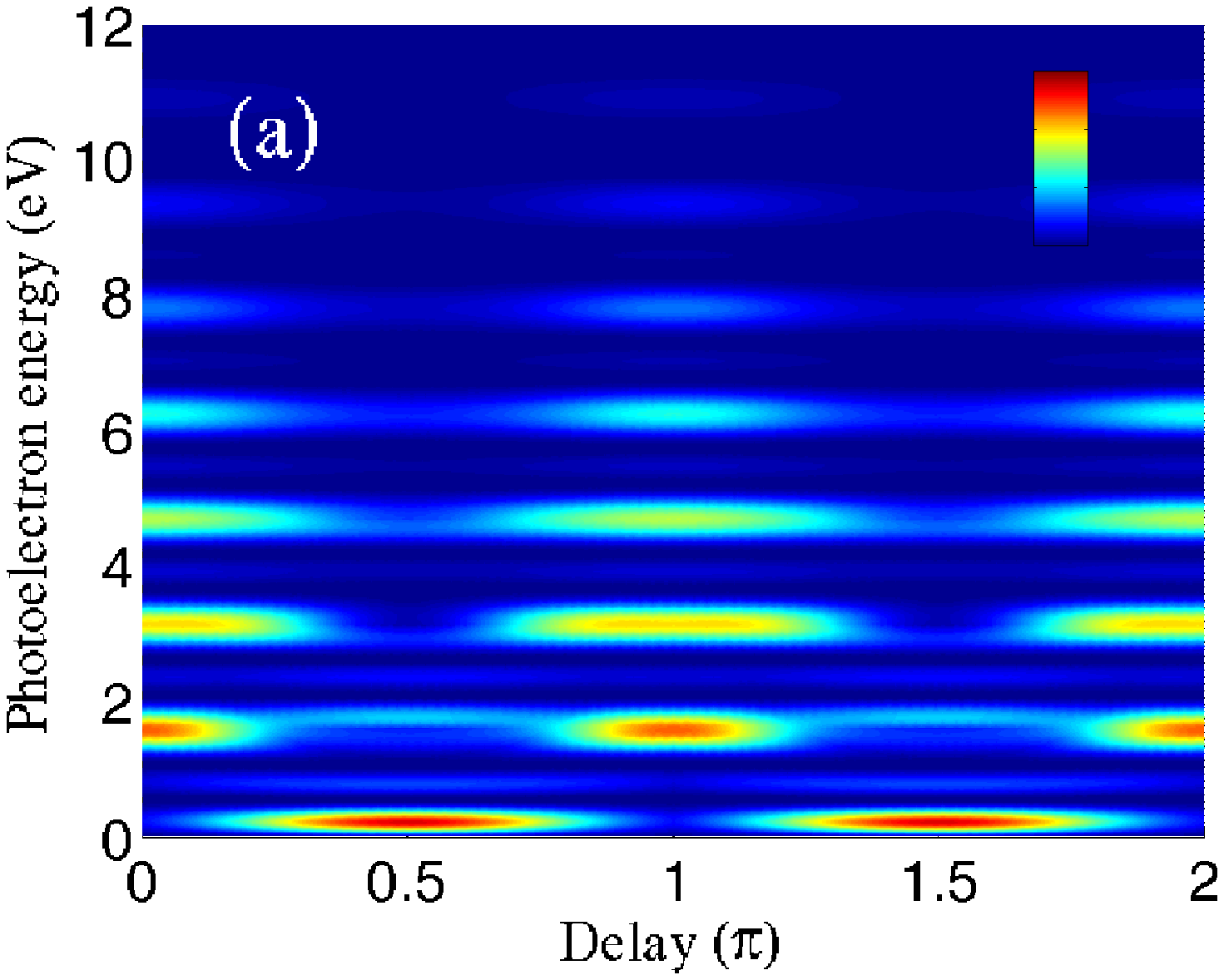}
\includegraphics[width=0.4\textwidth]{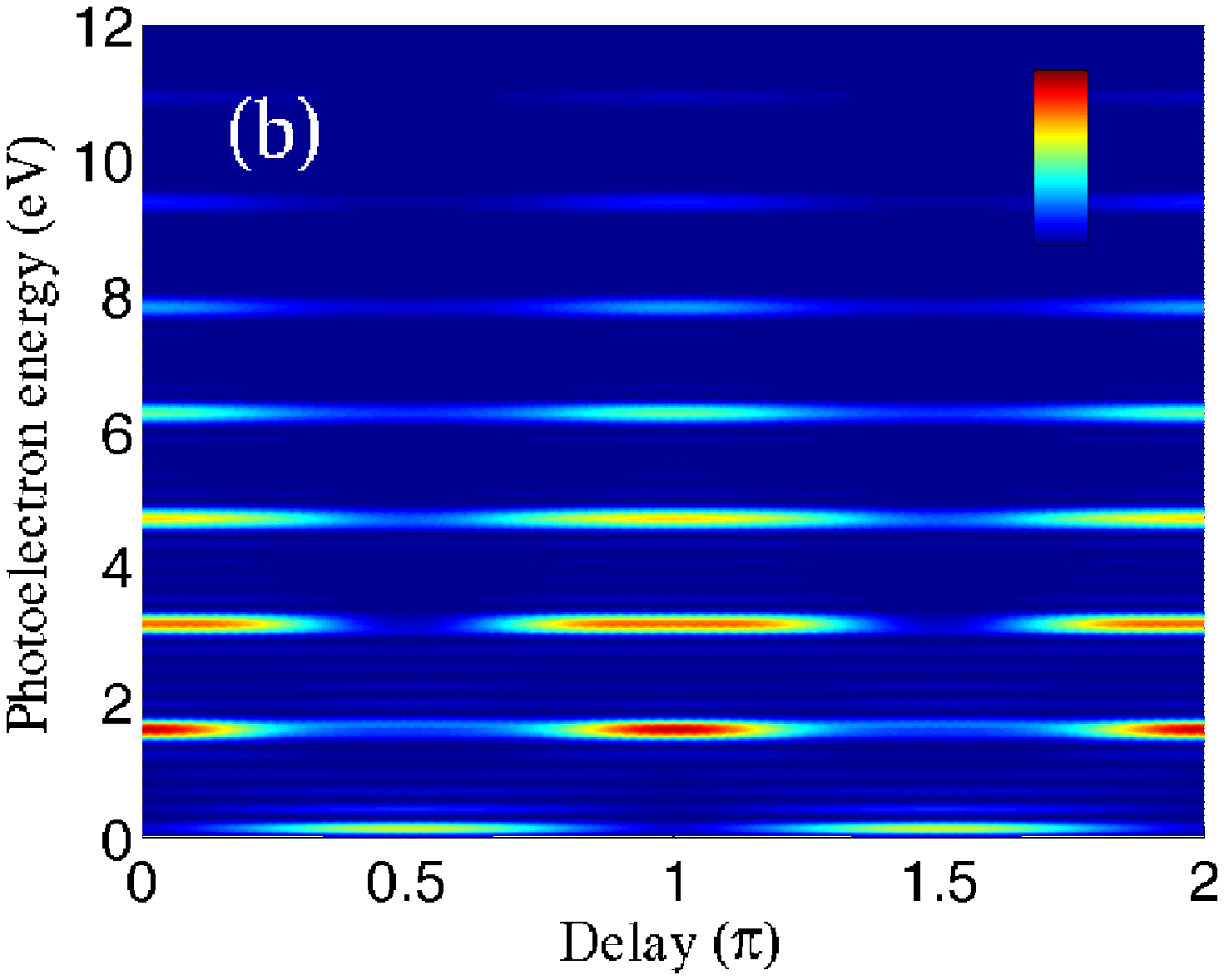}
\caption{Photo-electron energy spectra for an intensity $I=1.3\times 10^{13}$\,W/cm$^2$, excess energy $\epsilon_\f=-0.028$, $\nu =2$
and $N$=3 (a) and $N$=6 (b) IR cycles.}
\label{photoe}
\end{figure}
The effect of  an increasing number $N$  of IR cycles  in the comb $K_{N}(p)$ on the photo-electron spectrum
is shown in \fig{photoe} for the same intensity as before, but for 
an excess energy $\epsilon_\f=-0.028$ and $\nu = 2$. 
The narrower lines for larger $N$ (\fig{photoe}b as compared to \fig{photoe}a)
is due to the sharper comb for larger $N$.

For the cases shown in Figs.~\ref{fig:spectrum1},
\ref{fig:spectrum2} and \ref{photoe} the maxima appear at
$\varphi=0,\pi$ and the absorption 
probabilities have similar shapes. This is not always the case. Rather, 
the position of the maxima and the 
contrast between maxima and minima depend on 
the particular comb $K_{N}$ and therefore on $N$, $U_\mrm{p}$ and $\epsilon_\f$, as well as  on the details of the branches 
in the wave packet ${\cal X}_{\nu}$. This will be discussed in the next section.

%%%%%%%%%%%%%%%%%%%%%%%%%%%%%%%
\section{Position of the maxima and minima in the absorption probability}
We have seen how the oscillations in the absorption probability arise from the interplay between the comb function $K_{N}(p)$ and the momentum distribution ${\cal X}_{\nu}$ of a multi-component wave packet. Both
depend in a complex manner on the parameters of the laser fields and the groundstate energy of the atom. Hence the question arises if  one can predict analytically
for a given IR field intensity, where the
maxima  in $P_{\mrm{\nu}}(\varphi)$ appear as a function of phase delay for different VUV photon energies $\Omega$ of the APT.

\begin{figure}
\centering
\includegraphics[width=0.35\textwidth,height=4.5cm]{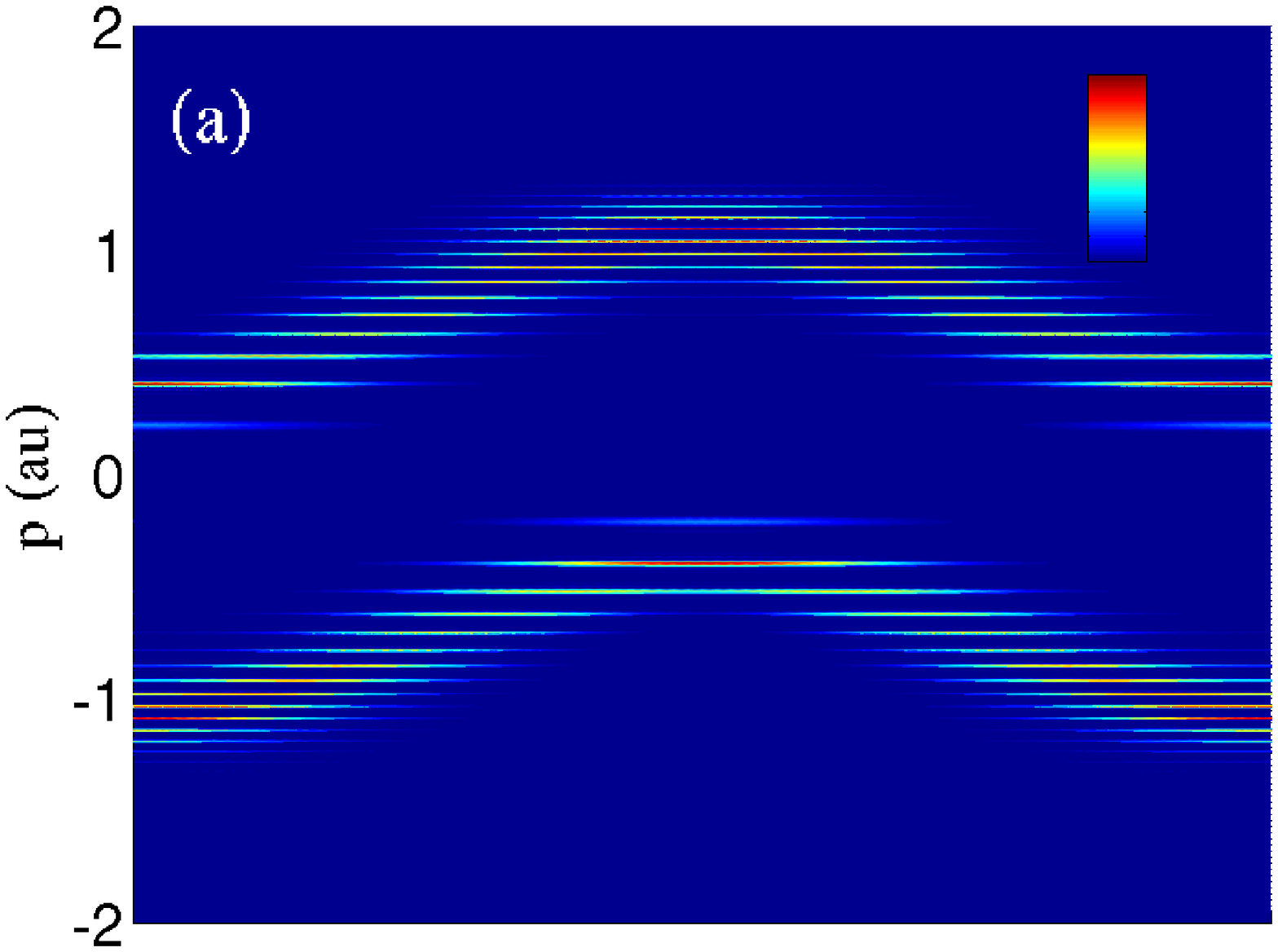}
\includegraphics[width=0.36\textwidth,height=4.4cm]{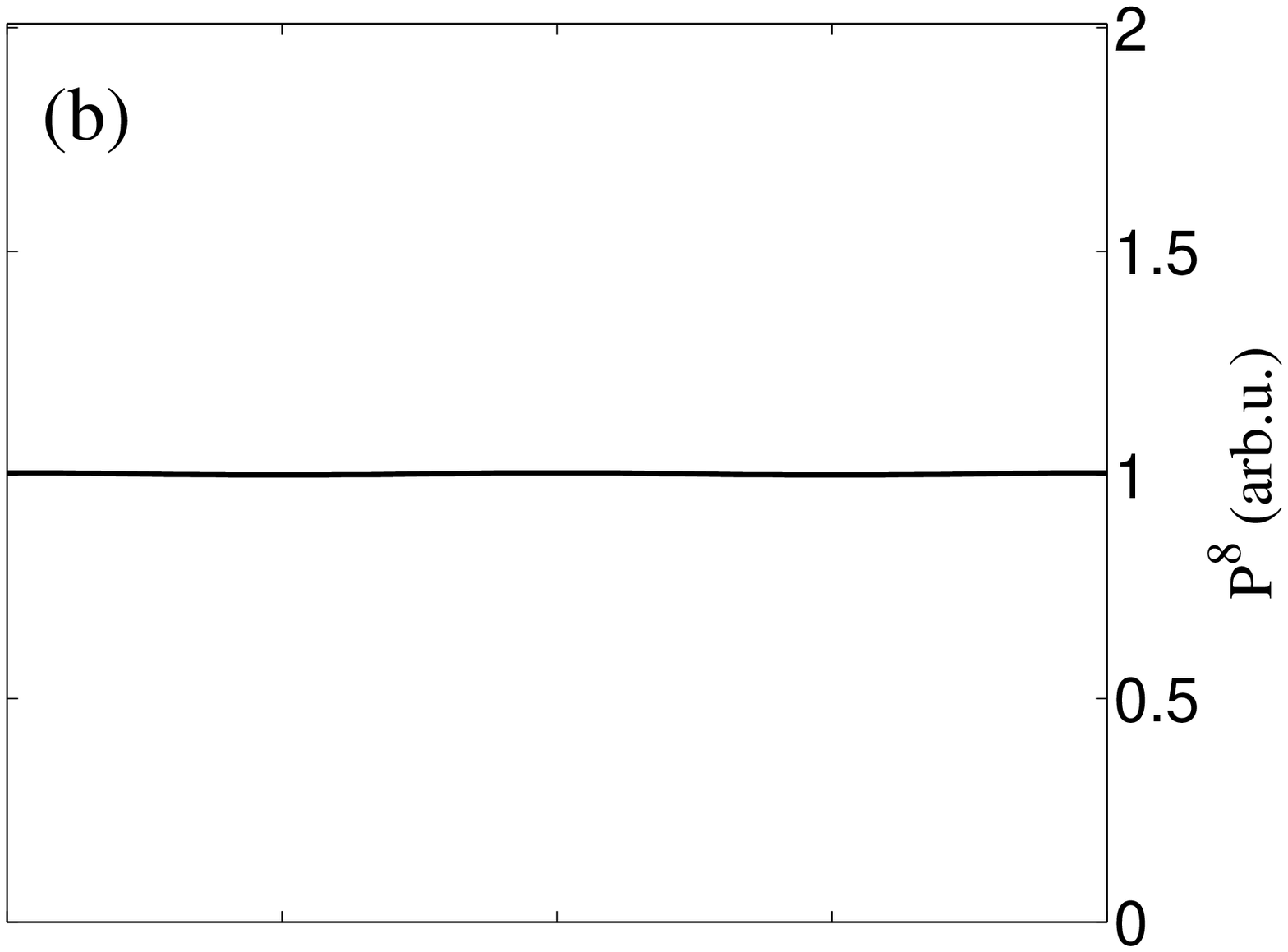}
\includegraphics[width=0.35\textwidth,height=4.6cm]{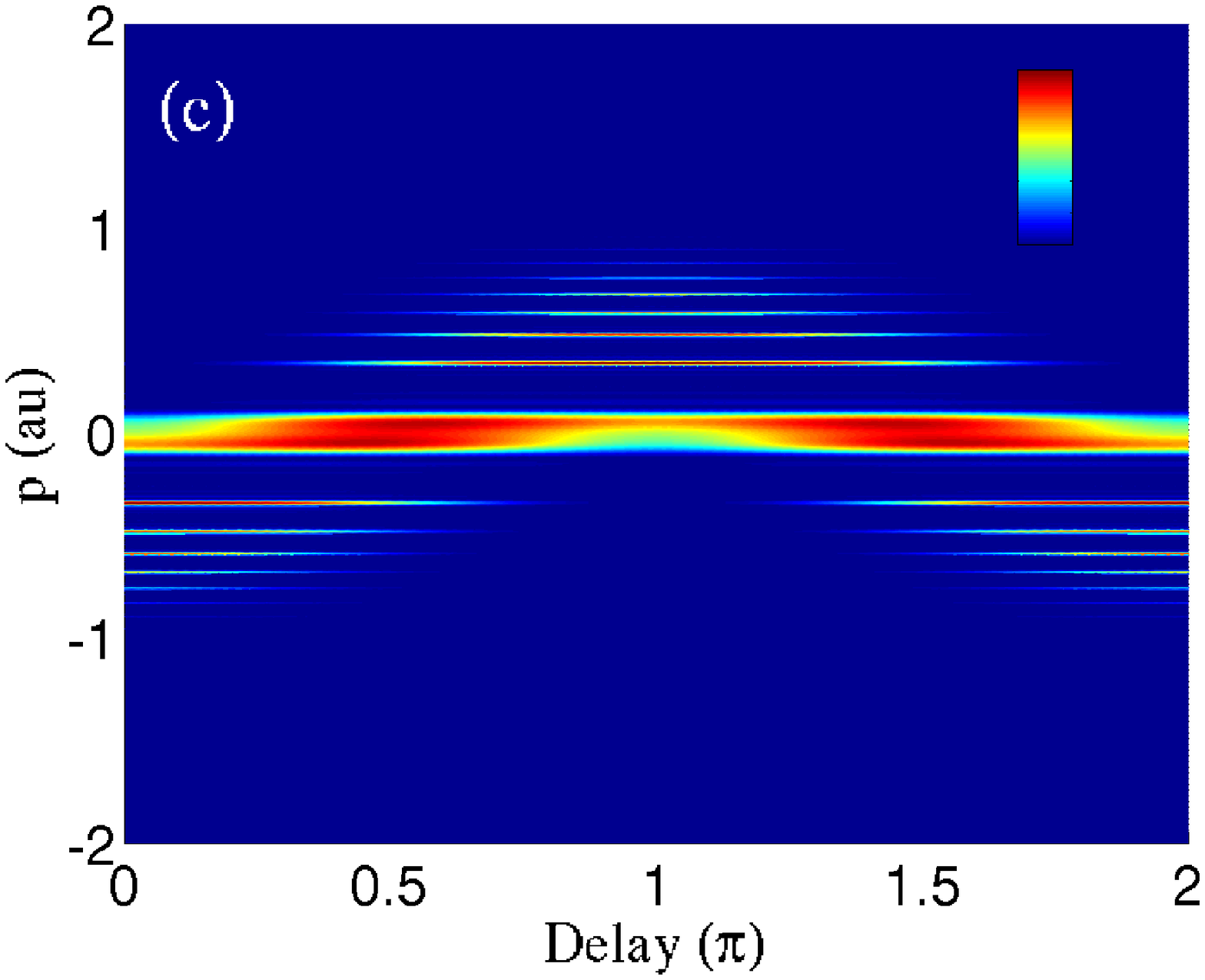}
\includegraphics[width=0.35\textwidth,height=4.4cm]{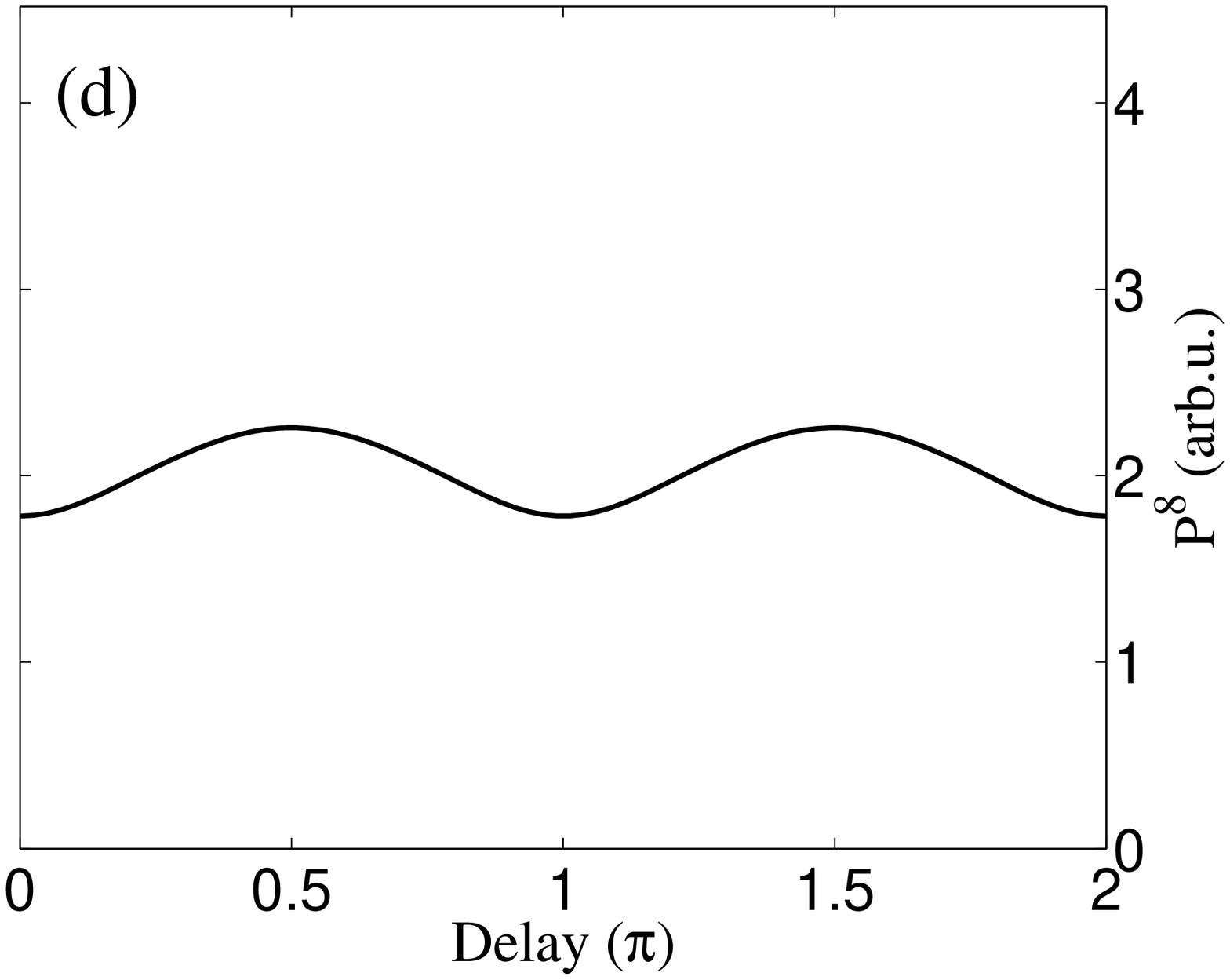}
\caption{Photo-electron momentum distribution (left) and
 corresponding absorption probability (right) for $\nu = 1$,
 $I=1.3\times 10^{13}$\,W/cm$^2$, $N=8$  and
 $\epsilon_\f=0.28$\,a.u. (upper panels) and
 $\epsilon_\f=-0.028$\,a.u. (lower panels).}  
\label{1ab}
\end{figure}

From the structure of the comb as discussed in 
Sect.~\ref{seccontrib} one can directly conclude that  
the oscillations in $P_{\mrm{\nu}}(\varphi)$ will disappear for  increasing $\epsilon_\f =\epsilon_\i+\Omega$, as illustrated in \fig{1ab}a. 
For large excess energy $\epsilon_\f$ the 
branches of ${\cal X}$ are centered about high absolute momentum
values $|\pm(2\epsilon_\f)^{1/2}\pm p_{\varphi}|$, where the comb is dense. 
Hence, the comb traces ${\cal X}(p,\varphi)$  homogeneously for all $\varphi$ and
the absorption probability hardly depends on $\varphi$ (\fig{1ab}b).
The physical meaning of this is that when the electronic wave packet triggered by an atto pulse
leaves the nucleus with a high kinetic energy, the overlap with
the EWPs released by subsequent atto pulses vanishes,
which diminishes the intereference among the wave packets.

A more systematic analysis of the position of the maxima and minima in the
absorption probability
can be carried out analytically using symmetry properties of $P_{\nu}$ and eventually a stationary phase
approximation with respect to $p$. 
The condition for extrema is $dP_{\nu}/d\varphi=0$, which can be written
for the case of $\nu=1$ as
\be{deriv1}
\frac{dP_1}{d\varphi}=\int\,dp K_N(p) {\cal X}'_1(p,\varphi)=0,
\end{equation}
where ${\cal X}'_1(p,\varphi) =\partial{\cal X}_{1}(p,\varphi)/\partial\varphi$, which is proportional to $x_{\varphi}$. Therefore, the absorption probability has extrema for $x_{\varphi}=0$. The symmetry of the functions under the integral reveals another set of maxima:
the comb $K_N(p)$ is even in $p$, and for $p_{\varphi}=0$,
the function ${\cal X}'_1$ is odd in $p$. Hence, the integral in \eq{deriv1}
is zero and there are also extrema for $p_{\varphi}=0$. To summarize, \eq{deriv1} is fulfilled 
for every $\varphi = n\pi/2$.

\begin{figure}
\centering
\includegraphics[width=0.4\textwidth]{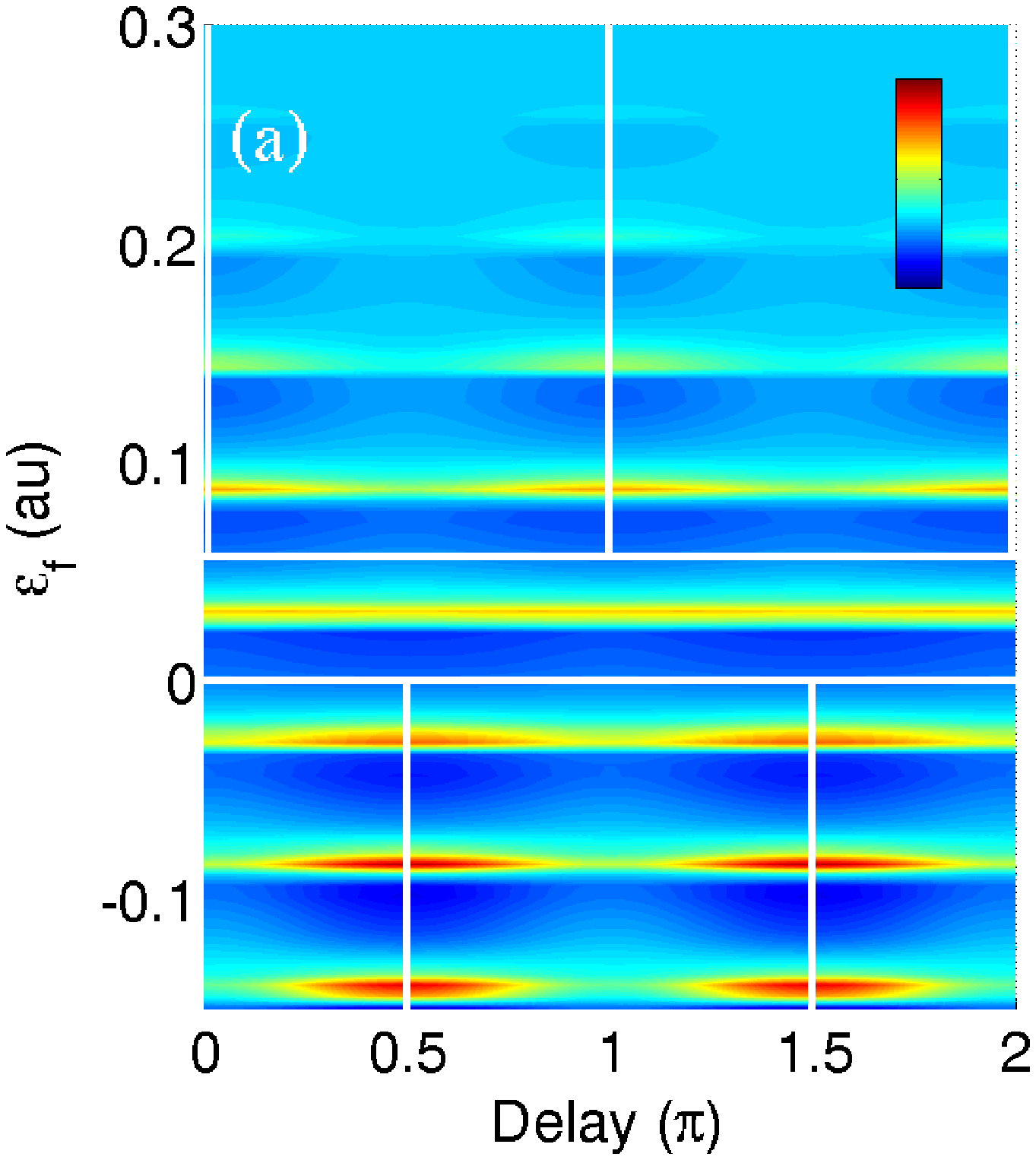}
\includegraphics[width=0.4\textwidth]{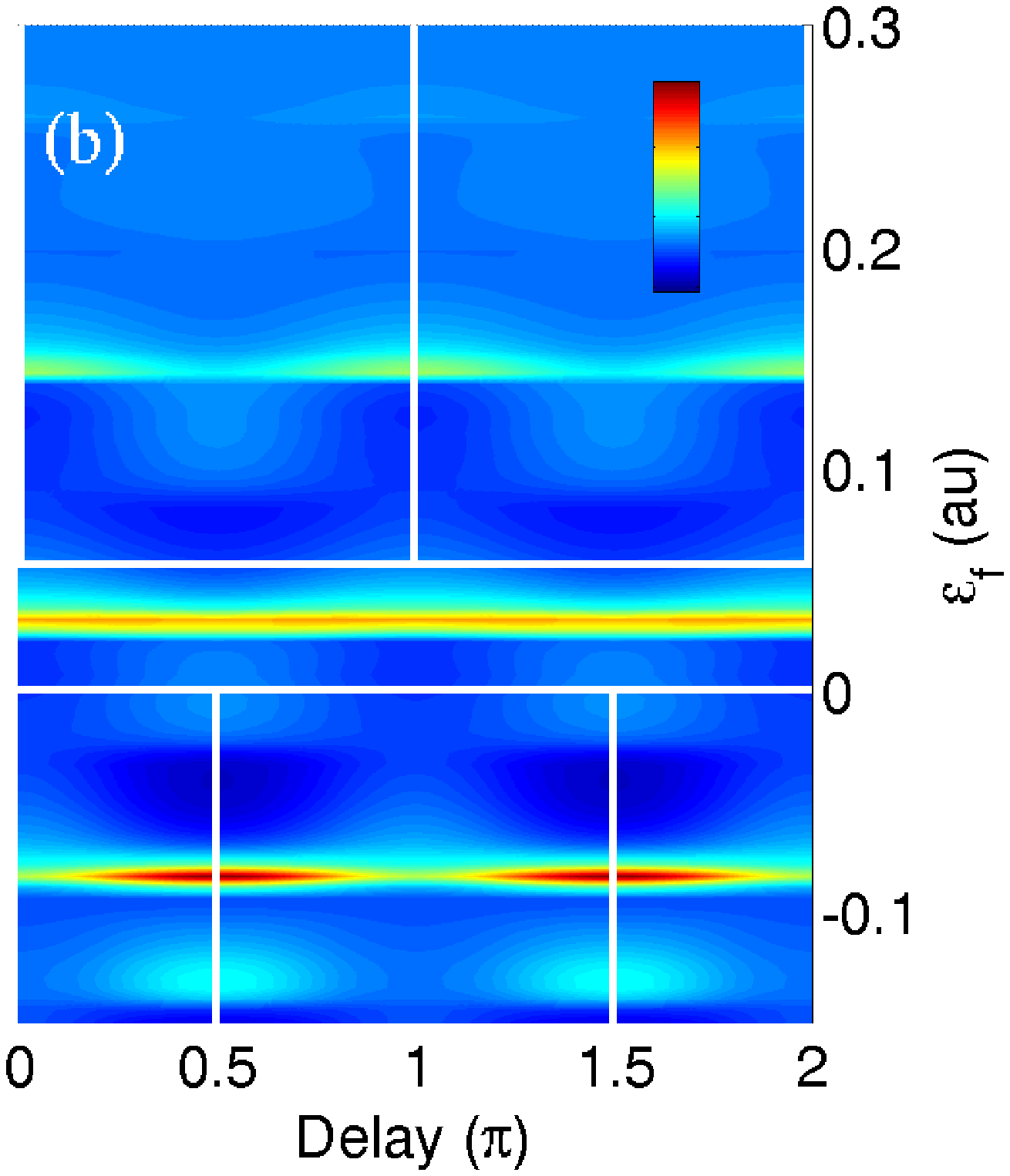}
\caption{Absorption probability as a function of $\varphi$, for 1 (a) and 2 (b) atto pulses per IR cycle,
for $I=1.3\times 10^{13}$\,W/cm$^2$ and $N$=8. 
Vertical white lines: position of the maxima at $\epsilon_\f$ in the two regions $\epsilon_\f<0$
and $\epsilon_\f>2U_\mrm{p}$, as predicted with the SPA: at $\varphi$=$\pi/2$ and $3\pi/2$ in the first case and 
at 0, $\pi$ and $2\pi$ in the second. The horizontal white lines differentiate these two regions.}
\label{carpet}
\end{figure}

To distinguish between maxima and minima we need the sign of the second derivative,
${\cal X}_{1}''$. To keep the derivation simple we will make now use of
the stationary phase approximation, which is applicable since  the comb $K_{N}(p)$ is a highly oscillatory function which depends only on $p^{2}$. Therefore, its global stationary phase point is   $p=0$, which is also obvious from \fig{fig:comb}, and we get  for \eq{deriv1}
\begin{equation}
\frac{dP_1}{d\varphi}\sim K_N(0) {\cal X}'_1(0,\varphi),
\end{equation}
where 
\begin{equation}\label{eq:chisSPA}
  {\cal X}'_1(0,\varphi)={\cal X}_1(0,\varphi)\;
  \frac{d}{d\varphi}\left[
    -\left(p_\varphi^2/2-\epsilon_\f\right)^2\Big/\sigma_{\mrm{\epsilon}}^2(0)\right]. 
\end{equation}
It can be easily shown that this function is zero for $x_\varphi=0$  or $p_\varphi=0$ as before, and
has additional zeros at $p_{\varphi}^{2}=2\epsilon_\f$. For the determination of maxima and minima we are only interested 
in the sign of the second derivative, which can be expressed for the three groups of extrema as
 \begin{eqnarray}\label{eq:signs}
  \mrm{sgn}\left[\left. {\cal X}_1''(0,\varphi)\right|_{\varphi = n\pi}\right]
  &=& \mrm{sgn}\left[g(2U_\mrm{p}-\epsilon_\f)\right]\\
  \mrm{sgn}\left[\left. {\cal X}_1''(0,\varphi)\right|_{\varphi = (n{+}1/2)\pi}\right]
  &=& \mrm{sgn}\left[g(\epsilon_\f)\right]
\end{eqnarray}
\begin{equation}
  \left.{\cal X}_1''(0,\varphi)\right|_{p_{\varphi}^{2}=2\epsilon_\f}
  = -8\frac{\epsilon_\f(2U_\mrm{p}{-}\epsilon_\f)}{\sigma_{\mrm{\epsilon}}^2} < 0
  \qquad 0\le\epsilon_\f\le 2U_\mrm{p}
\end{equation} 
with $g(x) = x(1+4\sigma^{4}\omega^{2}U_\mrm{p}x) 8\sigma^2 U_\mrm{p} {\cal X}_1(0,\varphi)$.
This leaves a clear structure of maxima and minima  for $\epsilon_\f<0$ and 
$\epsilon_\f>2U_\mrm{p}$ as summarized in Table~\ref{tabla}.
For $0<\epsilon_\f<2U_\mrm{p}$, all four extrema within $2\pi$ are minima, making the approximation not very trust worthy. Indeed, as will be demonstrated later, chaotic dynamics dominates this energy region rendering approximations problematic.
In \fig{carpet}a one can see this structure of maxima and minima
in $\varphi$ additionally modulated in $\epsilon_\f$, where the distance between the maxima is $\omega$. The latter is a consequence
of the effect of the comb function $K_{N}(0)$, which has maxima at $C_0\pi =
n\pi$ (with $C_0=C|_{p=0}$), 
which means maxima at energies $\epsilon_\f =U_\mrm{p}\pm n\omega$.

\begin{table}[!b]
\begin{center}
  \begin{tabular}{ | l | c | c|  }
  \hline
  energy  & $\varphi=n\pi$ & $\varphi=(n{+}1/2)\pi$ \\ \hline
  $\epsilon_\f>2U_\mrm{p}$  & maximum & minimum \\ \hline
  $\epsilon_\f<0$  & minimum & maximum \\ 
  \hline
  \end{tabular}
\end{center}
\caption{Position of the maxima and minima in $\varphi$
 for different energy ranges $\epsilon_\f$.}
\label{tabla}
\end{table}

Moving on to the case $\nu =2$ we investigate 
$P_{2}(\varphi)$ taking ${\cal X}_{2}$ from \eq{X2}, and for the sake of simplicity  
we will use the stationary phase approximation from the beginning. This function has the same stationary phase point
$p=0$ as before.  Then we can write 
%\begin{eqnarray}
%{\cal X}_{2}(0,\varphi)&=&2{\cal X}_{1}(0,\varphi)
%+2\cos(C\pi)\chi(0,\varphi)\chi^{*}(0,\varphi)\nonumber\\
%&=& 4\cos(C\pi/2){\cal X}_{1}(0,\varphi)\,.
%\label{X2app}\end{eqnarray}
\begin{equation}
  \label{X2app}
  {\cal X}_{2}(0,\varphi)=4\cos(C_0\pi/2){\cal X}_{1}(0,\varphi)\,.
\end{equation}
From \eq{X2app} it is immediately clear that the $\nu =2$ case has the same structure
of maxima and minima with respect to the phase delay $\varphi$ (see table \ref{tabla}) as the $\nu = 1$ case. The only difference is
the modulation in $\epsilon_\f$: The additional factor $\cos[C_0\pi/2]$ has maxima only
at even multiples of the IR frequency $\epsilon_\f=U_\mrm{p}\pm 2n\omega$, while  it has zeros
at odd multiples of $\omega$. Consequently, the distance between the peaks for $\nu =2$ on the $\epsilon_\f$ axis 
is given by $2\omega$ instead of $\omega$ as for $\nu =1$, which can be seen in \fig{carpet}.
%
%%%%%%%%%%%%%%%%%%%%%%%%%%%%
\section{Comparison with experimental results and exact quantum calculations}
\begin{figure}
\centering
\includegraphics[width=0.45\textwidth]{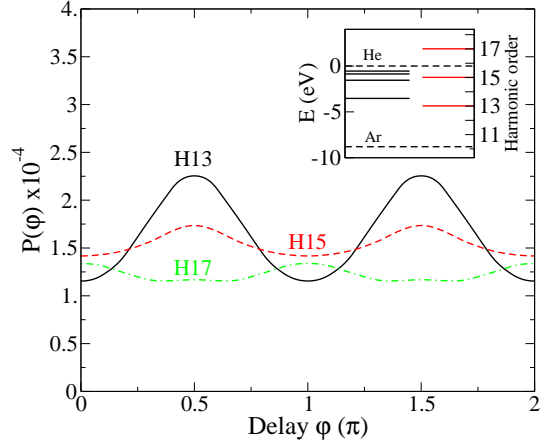}
\caption{Quantum calculation for the absorption probability for an IR with $\lambda$=796\,nm, $I=1.3\times 10^{13}$\,W/cm$^2$ and an APT with FWHM=5\,fs,
and three different central energies: the 13$^\mrm{th}$ harmonic of the IR field (full line), the 15$^\mrm{th}$ (dashed line) and the 17$^\mrm{th}$ 
(dashed-dotted line). 
The energy levels of the He atom are shown in the inset (full black lines), together with the ionization threshold of He and Ar (dashed lines).
The energies for the 13$^\mrm{th}$, 15$^\mrm{th}$ and 17$^\mrm{th}$ are also shown (red lines at the right).}
\label{piulf}
\end{figure}

The interesting dependence of the absorption probability on the
phase delay $\varphi$ was first reported experimentally and
shown to be in agreement with a full numerical quantum
calculation by Johnsson et al.\ \cite{Johnsson2007}. 
In the meantime it has been confirmed by other experiments
\cite{Cocke09}. 
Instead of increasing the VUV photon energy (to vary
$\epsilon_\f = \epsilon_\i+\Omega$), the ionization potential
$-\epsilon_\i$ was varied in the first experiment
\cite{Johnsson2007} by using He and Ar 
atoms as targets for  the combined IR + APT field, the latter with two atto pulses per IR cycle and a central energy of $\Omega$=23\,eV. This energy
is enough to ionize an electron from Ar, but not from He (see inset in \fig{piulf}). Strong oscillations in the ionization
probability of He as a function of $\varphi$  were found with maxima at $\varphi=(n+1/2)\pi$   and no oscillations were detected for Ar. 

These results are in qualitative agreement with our analytical predictions, as shown in \fig{carpet}b. 
For $\epsilon_\f<0$  (as in He with $\Omega$=23\,eV), we expect
maxima at $\varphi=(n+1/2)\pi$ , while for energies
well above threshold as in Ar, we expect a flat absorption
probability.  To double check the transition from the positions
of the maxima from $(n{+}1/2)\pi$ to $n\pi$ going from
$\epsilon_\f<0$ to positive $\epsilon_\f$, we have performed
full numerical calculations. We use a three-dimensional
one-electron model for the He atom\footnote{We propagated the TDSE in the atomic potential
$V(r)=-\left[1+\exp(-r/r_0)\right]/r$, with $r_0=1.05$\,{\AA}
guaranteeing the correct ionization potential of helium, and the
combined laser field of APT and IR pulse. The envelope of the
APT was a Gaussian of 5\,fs width, the one of the IR pulse had a 
$\cos^2$ shape containing 20 cycles.}, and a classical
electric field, with $\omega=0.0572$, $I=1.3\times 10^{13}$\,W/cm$^2$ 
for the IR, and $I=10^{11}$\,W/cm$^2$ for the APT.
Results are shown in \fig{piulf} for three different APTs, centered at the harmonics 13$^\mrm{th}$, 15$^\mrm{th}$ and 17$^\mrm{th}$, respectively. One sees indeed that the contrast of the maxima gets
smaller for increasing  but still negative $\epsilon_\f$, as it is the case going from the 
13$^\mrm{th}$ to the 15$^\mrm{th}$ harmonic  while for positive $\epsilon_\f$ (17$^\mrm{th}$ harmonic), there appear maxima at $\varphi=0,\pi$.

\begin{figure}
\centering
\includegraphics[width=0.45\textwidth]{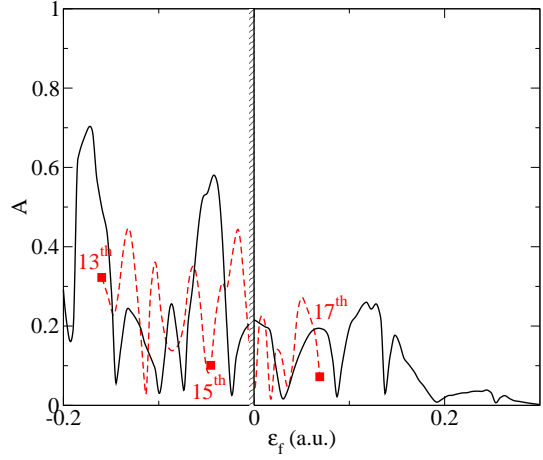}
\caption{Contrast (\eq{contrast})  of maxima and minima in the absorption probability for $\nu =2$ atto
pulses per IR period. Straight line: analytical approach. Dashed line: TDSE calculations,
for an APT with central energy from the 13$^\mrm{th}$ to 17$^\mrm{th}$  harmonics.
The energies for the 13$^\mrm{th}$, 15$^\mrm{th}$ and 17$^\mrm{th}$ harmonics are pointed out.
The conditions are $N=4$ ($N_{\mrm{atto}}=8$), $I=1.3\times 10^{13}$\,W/cm$^2$ 
and $\lambda=$796\,nm. The ionization threshold is marked by a vertical line.}
\label{cont}
\end{figure}

Hence, the simple man's approach presented here provides in general a very good understanding and interpretation of the effect a combined APT and IR field has on the ionization of atoms. Only for energies $\epsilon_\f$ very close to the ionization threshold,  $|\epsilon_\f|\approx U_\mrm{p}$
the simple man's approach is too drastic for reliable results.
In this energy range, the oscillatory absorption probability depends sensitively on details of the electronic wave packet 
whose dynamics is highly chaotic. This is demonstrated
in \fig{cont} for a sensitive observable, the  contrast between
maxima (at $\varphi_{\mrm{max}}$) and minima (at $\varphi_{\mrm{min}}$) for different photon energies,
\be{contrast}
A=\frac{P(\varphi_{\mrm{max}})-P(\varphi_{\mrm{min}})}{P(\varphi_{\mrm{max}})+P(\varphi_{\mrm{min}})}\,.
\end{equation}
One may question if such details of the chaotic behavior are helpful to understand the dynamics.
Future work will show if more robust observables, such as correlation functions with  characteristic correlation lengths and similar quantities  are more suitable to characterize electron dynamics under the illumination of APTs and IR fields.

%%%%%%%%%%%%%%%%%%%%%%%%%%%%%%%%%%%
\section{Conclusions and outlook}
We have  presented a minimal analytical approach to understand the behavior of electron wave packets generated by an 
attosecond pulse train (APT) in the presence of a strong IR field in the framework 
of the simple man's approach in strong field physics, with special emphasis in the phase delay between the APT and IR pulses. 
In this approximation the photo absorption probability can be written as a product of a frequency comb function, resulting from the periodic 
nature of the pulses with the IR frequency, and the probability density of an electronic wave packet
whose number of components is given by the number of attosecond pulses within on IR period.
Only the latter depends on the phase delay between APT and IR field, while the former acts like an electron momentum filter. The minimal approach provides insight into the formation of oscillations in the photo-absorption as a function of phase delay, on the frequency of these oscillations  and the general trend of the phase delay as a function of excess (or photon) energy. The minimal approach fails for excess energies comparable with the ponderomotive potential,
where  the chaotic nature of the electron dynamics renders results very sensitive to approximations.

\appendix\section[\mbox{}\hspace{4em}Taylor expansion of the
phase]{Taylor expansion of the phase}\label{sec:TaylCoeff}
%\vspace{0.25in}\noindent{\bfseries Appendix}\vspace{-0.25in}\nopagebreak
%\appendix\def\thesection{\Alph{section}}
%\section{Taylor expansion of the phase}\label{sec:TaylCoeff}
The functions $F^{(k)}_n=d^{k}\phi_n(t)/dt^{k}\Big|_{t=t_n}$ needed in \eq{uninf} are 
\begin{eqnarray}
F^{(0)}_n=i\int^{t_n}\frac{(p+A(\tau))^2}{2}\,d\tau-i\epsilon_\f t_n\,,  \\
F^{(1)}_n=i\frac{(p+A(t_n))^2}{2}-i\epsilon_\f\,, \\
F^{(2)}_n=i(p+A(t_n))A'(t_n)-\frac{1}{\sigma^2}\,.
\end{eqnarray}
Their explicit values for $t_n=2\pi n/\omega$ are
\begin{eqnarray}
  \label{f0}F^{(0)}_n=F_0+i2n\pi C +ipx_{\mrm{\varphi}}\,,\\
  \label{f1}F^{(1)}_n=i\frac{(p+p_{\mrm{\varphi}})^2}{2}-i\epsilon_\f\,,\\
  \label{f2}F^{(2)}_n=-i(p+p_{\mrm{\varphi}})\omega^2x_{\mrm{\varphi}} -\frac{1}{\sigma^2}\,,
\end{eqnarray}
where $C$, $x_{\varphi}$ and $p_{\varphi}$ are defined in
\eq{Cgl} and \eq{quiver}, respectively.
$F_0$ is a phase that does not contribute to $|\psi^{\infty}_{\nu}(p,\varphi)|^2$ (\eq{dpdp}).

For two atto pulses per IR cycle ($\nu=2$) we have in additional
set of quantities $F^{(k)}_{n{+}1/2}$.
They differ from the values of the $F^{(k)}_n$ in
Eqs.~(\ref{f0})\,--\,(\ref{f2}) only in two respects:
The term $i2n\pi C$ in $F^{(0)}_n$ has to be modified to
$i(2n{+}1)\pi C$ and in all quantities $p$ has to be replaced by
$-p$, which follows from of the Volkov propagator (\ref{volkov}).

%%%%%%%%%%%%%%%%%%%%%%%%%%%%%%%%%%%%%%%%%%%%%%%%%%%%%%%%%%%%

\end{document}